%Paper: astro-ph/9404024
%From: mrc@namaste.UCSC.EDU (Marc Herant)
%Date: Tue, 12 Apr 1994 11:02:13 -0700

%plain tex
\magnification 1100
\baselineskip=15truept
\parskip=\smallskipamount

\centerline{\bf INSIDE THE SUPERNOVA:}
\medskip
\centerline{\bf A POWERFUL CONVECTIVE ENGINE}
\bigskip
\centerline{\bf Marc Herant$^{1,2}$, Willy Benz$^3$}

\centerline{\bf W. Raphael Hix$^4$,
Chris L. Fryer$^3$, and Stirling A. Colgate$^5$}
\bigskip
\centerline{$^1$Board of Studies in Astronomy and Astrophysics}
\centerline{University of California, Santa Cruz}
\centerline{Santa Cruz, CA 95064}
\centerline{mrc@lick.ucsc.edu}
\medskip
\centerline{$^2$ Supported by a {\it Compton Gamma Ray Observatory} Fellowship}
\medskip
\centerline{$^3$Steward Observatory}
\centerline{University of Arizona}
\centerline{Tucson, AZ 85721}
\medskip
\centerline{$^4$Harvard-Smithsonian Center for Astrophysics}
\centerline{60 Garden St.}
\centerline{Cambridge, MA 02138}
\medskip
\centerline{$^5$Theory Division}
\centerline{Los Alamos National Laboratory, MS B275}
\centerline{Los Alamos, NM 87545}
\bigskip
\centerline{Submitted to:}
\centerline{\it The Astrophysical Journal}
\centerline{December 1993}
\centerline{Revised:}
\centerline{March 1994}
\bigskip
\centerline{ABSTRACT}
\bigskip

We present an extensive study of the inception of supernova
explosions by following the evolution of the cores of two
massive stars (15 M$_\odot$ and 25 M$_\odot$) in multidimension.
Our calculations begin at the onset of core collapse and stop
several hundred milliseconds after the bounce, at which time
successful explosions of the appropriate magnitude have been
obtained. Similar to the classical delayed explosion mechanism of
Wilson (1985), the explosion is powered by the heating of the
envelope due to neutrinos emitted by the protoneutron star as
it radiates the gravitational energy liberated by the collapse.
However, as was shown by Herant, Benz \& Colgate (1992), this
heating generates strong convection outside the neutrinosphere,
which we demonstrate to be critical to the explosion.  By
breaking a purely stratified hydrostatic equilibrium, convection
moves the nascent supernova away from a delicate radiative
equilibrium between neutrino emission and absorption.  Thus, unlike
what has been observed in one-dimensional calculations, explosions
are rendered quite insensitive to the details of the physical input
parameters such as neutrino cross-sections or nuclear equation
of state parameters. As a confirmation, our comparative
one-dimensional calculations with identical microphysics, but
in which convection cannot occur, lead to dramatic failures.

Guided by our numerical results, we have developed a paradigm for
the supernova explosion mechanism. We view a supernova as an open
cycle thermodynamic engine in which a reservoir of low-entropy
matter (the envelope) is thermally coupled and physically
connected to a hot bath (the protoneutron star) by a neutrino
flux, and by hydrodynamic instabilities. Neutrino heating raises
the entropy of matter in the vicinity of the protoneutron star until
buoyancy carries it to low density, low temperature regions at
larger radii.  This matter is replaced by low-entropy downflows with
negative buoyancy.  In essence, a Carnot cycle is established in
which convection allows out-of-equilibrium heat transfer mediated
by neutrinos to drive low entropy matter to higher entropy and
therefore extracts mechanical energy from the heat generated
by gravitational collapse.

We argue that supernova explosions are nearly guaranteed and
self-regulated by the high efficiency of the thermodynamical
engine.  The mechanical efficiency is high because mixing during the
heat exchange is limited by the rapid rise and shape-preserving
expansion of the bubbles in a $\rho\propto r^{-3}$ atmosphere. In
addition, the ideal Carnot efficiency is high due to the large
temperature contrast between the surface of the protoneutron
star and the material being convected down from large radii
(this contrast remains large in spite of compression and shock
heating which is relatively small).  By direct $P\,dV$ integration
over the convective cycle, we estimate the energy deposition
to be $\sim4$ foes per M$_\odot$ involved. Further, convection,
by keeping the temperature low in rising neutrino-heated
high-entropy bubbles, allows the storage of internal energy
while minimizing the losses due to neutrino emission.  Thus
convection continues to accumulate energy exterior to the neutron
star until a successful explosion has occurred.  At this time,
the envelope is expelled and therefore uncoupled from the heat
source (the neutron star) and the energy input ceases.  This
paradigm does not invoke new or modified physics over previous
treatments, but relies on compellingly straightforward
thermodynamic arguments. It provides a robust and self-regulated
explosion mechanism to power supernovae which is effective
under a wide range of physical parameters.

\medskip\noindent
{\it Subject headings:} hydrodynamics -- stars: interiors --
stars: neutron -- supernovae: general

\vfill\eject
\end